\documentclass{osa-article}
\journal{oe}

\articletype{Research Article}
\DeclareMathOperator*{\argmin}{arg\,min}
\begin{document}

\title{Ultra-high resolution and broadband chip-scale speckle enhanced Fourier-transform spectrometer}

\author{Uttam Paudel,\authormark{1,*} Todd Rose,\authormark{1}}

\address{\authormark{1}Electronics and Photonics Laboratory, The Aerospace Corporation, 2310 E El Segundo Blvd, El Segundo, CA, 90245}

\email{\authormark{*}uttam.paudel@aero.org} 

\begin{abstract}
Recent advancements in silicon photonics are enabling the development of chip-scale photonics devices for sensing and signal processing applications, among which on-chip spectrometers are of particular interest for precision wavelength monitoring and related applications. Most chip-scale spectrometers suffer from a resolution-bandwidth trade-off, thus limiting the uses of the device. Here we report on a novel passive, chip-scale, hybrid speckle-enhanced Fourier transform device that exhibits a two order-of-magnitude improvement in finesse (bandwidth/resolution) over the state-of-the art chip-scale speckle and Fourier transform spectrometers. In our proof-of-principle device, we demonstrate a spectral resolution of 140 MHz with 12-nm bandwidth for a finesse of $10^4$ that can operate over a range of 1500-1600 nm. This chip-scale spectrometer structure implements a typical spatial heterodyne discrete Fourier transform interferometer network that is enhanced by speckle generated from the wafer substrate. This latter effect, which is extremely simple to invoke, superimposes the high wavelength resolution intrinsic to speckle generated from a strongly guiding waveguide with a more broadband but lower resolution discrete Fourier transform modality of the overarching waveguide structure. This hybrid approach signifies a new pathway for realizing chip-scale spectrometers capable of ultra-high resolution and broadband performance. 
\end{abstract}

\section{Introduction}
Chip-scale optical spectrometers are envisioned as key elements for next generation remote sensing systems and precision on-chip wavelength monitoring. A primary example is the development of optical devices for detection of chemical species on remote platforms\cite{scott_2004, cui_2012, subramanian_2015, kita_review_2017, lin_2017, holmstrom_2016}. For space applications, a compact, high-resolution, and alignment-free spectrometer with no moving parts or exposed surfaces is highly desirable. Recent advances in silicon photonics are enabling the fabrication of such devices using CMOS compatible commercial foundries and offer significant size, weight, and power (SWAP) along with cost advantages over traditional spectrometers. 

Silicon-on-insulator (SOI) discrete Fourier transform (DFT) \cite{florjanczyk_2007, okamoto_2010, velasco_2013, nedeljkovic_2016, kita_2018} and multimode waveguide (MMW)-based speckle spectrometers \cite{redding_2012, redding_2013, reddingNatPhot_2013, piels_2017, scofield_2018} are two promising architectures. Despite their versatile performance, both technologies suffer from a bandwidth-resolution tradeoff \cite{okamoto_2010, velasco_2013, nedeljkovic_2016, redding_2014}. For example, the spectral resolution ($\delta\lambda$) of a DFT spectrometer is inversely proportional to the maximum path-length difference ($\Delta L_{max}$) of the interferometers used, and the bandwidth ($\Delta\lambda$) and the finesse ($\Delta\lambda/\delta\lambda = N/2$) are set by the total number of interferometers (N). Similarly, the spectral resolution of a speckle spectrometer derived from a MMW (both chip-scale and with a multimode fiber) is inversely proportional to its length, and its bandwidth is given by $\Delta\lambda\sim (N_m-1)\delta\lambda$, where the finesse is set by the number of distinct speckle modes ($N_m$) available for the measurement \cite{redding_2014}. As constructing a single spectrometer with a large finesse will require fabricating thousands of interferometers or spectral multiplexing or other exotic configurations, the finesses of the current state-of-the art chip-scale DFT and MMW speckle devices are limited to few hundred \cite{florjanczyk_2007, okamoto_2010, velasco_2013, nedeljkovic_2016, kita_2018,redding_2012, redding_2013, reddingNatPhot_2013, piels_2017, scofield_2018}; thus, a larger finesse device would be highly desirable.  

In this Article, we report on a first demonstration of a speckle-enhanced DFT (SDFT) chip-scale spectrometer which combines both modalities on a single chip to achieve a finesse that exceeds the current state-of-the-art performance of either chip-scale spectrometer types by two orders of magnitude. This combination of DFT and speckle modalities yields a device with the broader bandwidth characteristics of the DFT and the higher resolution characteristics of the MMW without increasing the number of structures or increasing the size of the device.  

\section{Spectrometer design and operation}
\subsection{Device layout}

Traditional Fourier-transform spectrometers incorporate an unbalanced Michelson interferometer where light intensity is monitored at an output port while scanning the relative path length difference between the two arms. By taking the Fourier transformation of the recorded intensity, the unknown input spectrum is reconstructed. An analogous discrete Fourier transform spectrometer, also referred to as a spatial heterodyne FT spectrometer, implements an array of unbalanced Mach-Zehnder interferometers with predefined relative path length differences \cite{florjanczyk_2007, okamoto_2010, velasco_2013, nedeljkovic_2016}. In this manner, no active scanning mechanism or alignment procedures are required, making them more robust. Data throughput is potentially improved due to parallel processing needed for this approach.  

\begin{figure}[h!]
\centering\includegraphics[width=8cm]{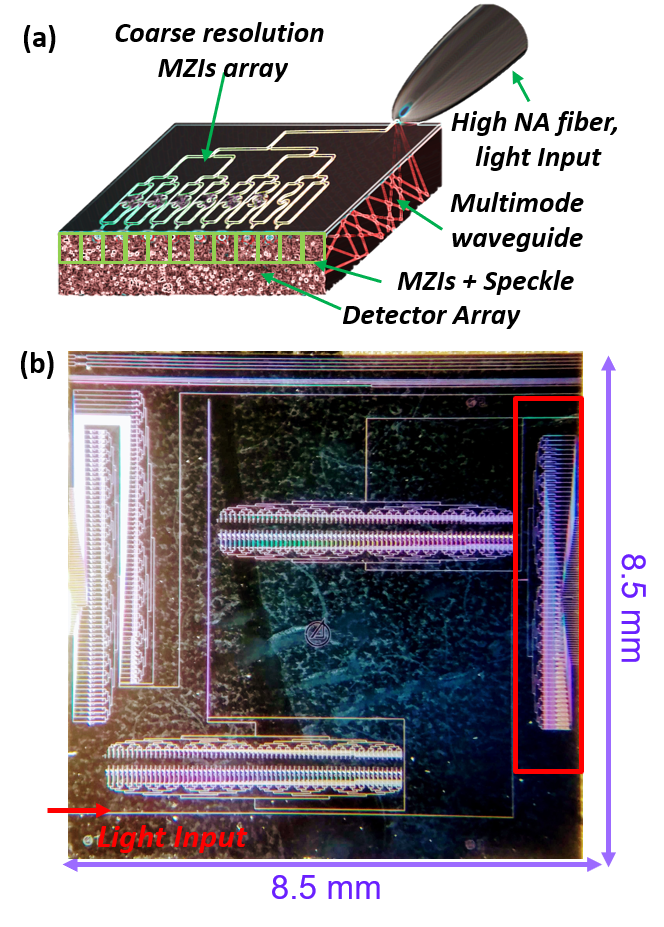}
\caption{(a) A schematic diagram of the hybrid speckle and discrete Fourier transform (SDFT) spectrometer consisting of a thick wafer substrate that acts as a multimode waveguide and a series of MZIs with varying relative path length differences fabricated on top of it. Light is coupled to the chip through one end of the chip using a single-mode fiber and is imaged at the other end using an array detector.(b) Microscope image of fabricated die consisting of 4 different spectrometers with 64 and 128 MZIs arrays. The device used in this study is enclosed by the red box.}
\label{fig:figure1}
\end{figure}

The DFT functionality of our SDFT spectrometer is implemented using such spatial heterodyne interferometers on an SOI platform. The layout for the SDFT spectrometers was developed using an open-source PDK developed by SI-EPIC \cite{chrostowski_2016} and fabricated using a CMOS compatible commercial foundry (Applied Nanotools). The Mach-Zehnder interferometers (MZIs) were fabricated using single-etch electron-beam lithography on a 220 nm thick layer of silicon. Arrays of 64 and 128 MZIs were built using 500 nm wide single-mode waveguides (that sustain both $TE_{00}$ and $TM_{00}$ modes) with 50-$\mu$m stepwise incremental pathlength differences. A schematic diagram of the SDFT spectrometer along with a microscope image of the fabricated device are shown in Fig.\ref{fig:figure1}. The DFT functionality used in this study, which consists of 64 MZIs, is enclosed by the red box.

The MZIs in the array are optically coupled via a cascaded network of 3 dB Y-splitters. The output of the MZIs are terminated at the end of the chip where the light emission is imaged on a camera array. A 2 $\mu$m thick oxide layer is grown on top of the interferometers for protection and to reduce thermal sensitivity. The optical input waveguide is off-centered from the field of view of the camera imaging the output waveguides to minimize any leaked light propagating on top of the device that otherwise would saturate the detector arrays. 

The speckle functionality is derived from the 8.5 mm long 675 $\mu$m thick silicon handler wafer which behaves as a strongly guiding planar MMW that sits below the DFT waveguides. The silicon waveguide sustains several hundred thousand optical modes that yield highly developed speckle at the output of the chip.

Light is coupled to the MZIs and the MMW through one end of the chip using a high-NA single-mode fiber (NA=0.41) and the output of the chip is imaged using a 4x microscope objective and a high-speed InGaAs camera array (GoodRich SU640KTS). The input and output of the waveguides are terminated at the end of the chip-facet using a polarization independent, sub-wavelength grating mode-converter. The mode-converter uses an inverse-tapered structure to expand the 500 nm waveguide mode to a 3 $\mu$m mode field diameter in order to maximize the coupling with the high-NA fiber \cite{cheben_2015, steidle_2015}. Due to a slight mode mismatch between the fiber and the tapered waveguide structure, a fraction of the input light is leaked into the substrate MMW. This results in the simultaneous propagation of the optical beam through both the MZIs (DFT spectrometer) and the MMW (Speckle spectrometer), thus forming a hybrid SDFT spectrometer.

\subsection{Theory of operation}

The electric field output of the optical modes propagated through an ideal substrate waveguide with length L can be written as \cite{redding_2013}
\begin{equation}
E_{speckle} (x,y,\lambda) = \sum_m C_m \psi_m(x, y, \lambda) \mbox{exp}[-i(\beta_m(\lambda)L-\omega t+\phi_m)]]
\label{eq:refname1a}
\end{equation}

where $\psi_m$ is the spatial profile of the $m^{th}$ mode that has initial amplitude $C_m$ and phase $\phi_m$ with propagation constant $\beta_m$ and is measured at $(x,y)$ coordinate at the output facet of the waveguide. A large width slab waveguide sustains several thousand optical modes and the interference between those modes results in a speckle pattern. In addition, the propagation constant is wavelength dependent and as a consequence, any change in the input wavelength results in the modification of the output interference pattern that generates a unique wavelength-dependent fingerprint. Similarly, the electric field of the MZI array output is 

\begin{equation}
E_{MZI} (x, y,\lambda) = \frac{C}{2}\sum_{n=1}^N\psi_n(x, y,\lambda) \mbox{exp}[i\phi] (1+\mbox{exp}[-i \beta \Delta L_n])
\label{eq:refname2a}
\end{equation}

where $\psi_n(x, y,\lambda)$ is the spatial mode distribution at the output of the waveguide and $\Delta L_n$ is the relative path-length difference of nth MZI. 

The total output power recorded by a camera at the output facet of the chip can be calculated by taking the modulus square of the sum of the output electric fields from both the MMW and MZI array. The SDFT device can be effectively treated as a transformation operator ($A_{SDFT}(x,y,\lambda)$) that maps the input spectral information ($S(\lambda)$) to the output spatial dependent intensity pattern,   
\begin{equation}
P_{out}(x,y,\lambda)= A_{SDFT}(x,y,\lambda).S(\lambda)
\label{eq:refname3}
\end{equation}
where the transformation operator is the sum of wavelength-spatial responses from the component MZI arrays and the MMW substrate (speckle), and any cross term that could arise from the interference between modes from both components. An exact first principles calculation of such a transformation matrix is a complicated task; however, one can experimentally measure the wavelength-and spatial-dependent transmission matrix ($A_{SDFT}$), or calibration matrix, by tuning a narrow bandwidth single-frequency laser and recording the intensity pattern with a camera. 

\section{Results: device characterization}
\subsection{Calibration}

\begin{figure}[ht]
\centering
\includegraphics[width=13cm]{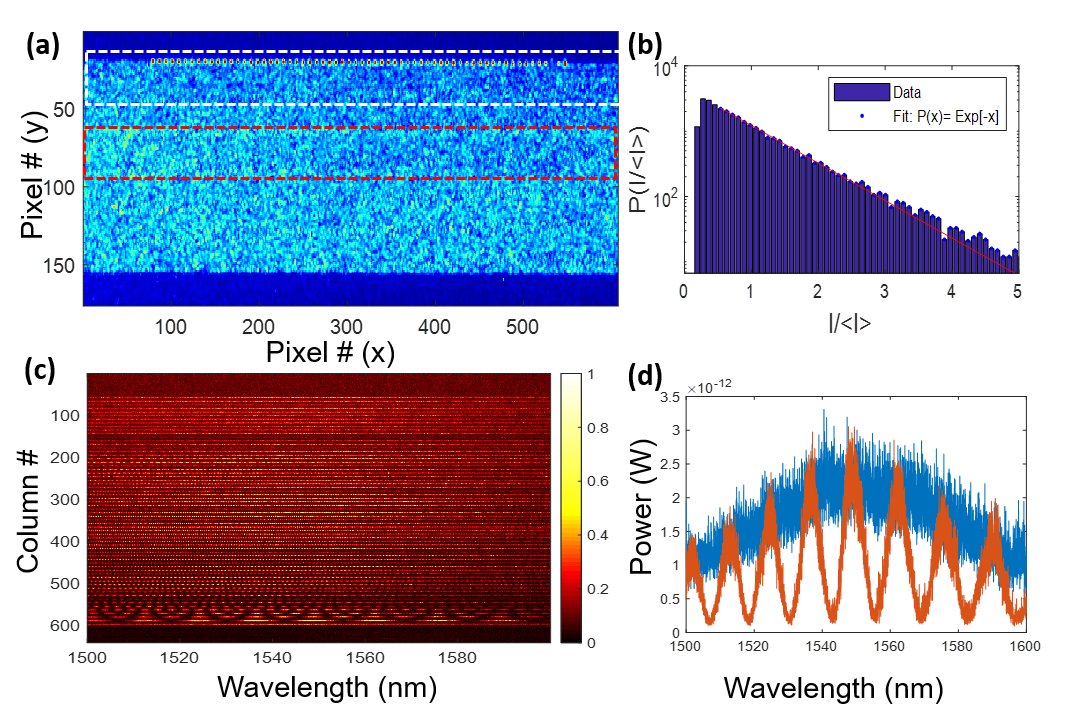}
\caption{(a) Recorded output image of a 64 MZI SDFT chip for monochromatic input light at 1550 nm. The dashed white box represents a SDFT region consisting of both MZI and speckle output and the red box consists of speckle-only output from the MMW. (b)  Intensity probability distribution function calculated from the intensity data recorded at the speckle-only region of the chip, fit with a negative exponential decay, where the x-axis is the intensity value normalized by the mean and the y-axis is the number of pixels within each intensity bin. Such negative exponential decay is evidence of a fully developed speckle pattern arising from the multimode interference of the beam propagating through the substrate waveguide \cite{goodman_1976}. (c) Wavelength dependent transmission matrix recorded from the output of the spectrometer recorded over 100 nm spectral window. The 2D camera image within the region of interest is summed column-wise to generate a 1D pixel array. (d) Transmission profile of a $\Delta L = 0$ $\mu$m and $50$ $\mu$m MZIs each recorded from a single element of the 1D array by tuning the wavelength over a 100 nm spectral window and consist of contributions from both speckle and MZI outputs.
}
\label{fig:figure2}
\end{figure}

Figure \ref{fig:figure2} (a) is an experimentally recorded image of the output of the SDFT chip. The dashed white box represents a SDFT region consisting of both MZI and speckle output and the red box consists of speckle-only contribution from the MMW. The intensity distribution of the output of the MMW recorded by the camera is plotted in Fig.\ref{fig:figure2}(b) along with a negative-exponential decay fit, where the x-axis is the intensity normalized by the mean and y-axis is its distribution. This negative-exponential decay in the intensity pattern is a characteristic of fully developed speckle resulting from the interference of a large number of modes \cite{goodman_1976}. 

The region of the chip corresponding to the SDFT and speckle-only output are summed column-wise to generate a 1D pixel array of the calibration matrix at a discrete wavelength step for the corresponding device. Figure \ref{fig:figure2} (c) is an example SDFT calibration matrix recorded from the output of a 64-element SDFT chip over a 100 nm spectral window generated by scanning a narrow band continuous-wave (CW) laser. For a wavelength dependent 2D intensity output of the SDFT device, see Visualization 1. Figure \ref{fig:figure2} (d) shows the transmission profiles of two MZIs with $\Delta L = 0$ $\mu$m and $50$ $\mu$m recorded over a 100 nm spectral window and consists of intensities from both MZIs and the MMW, where the speckle corrupts the MZIs' transmission as a high-frequency noise, forming a unique wavelength-dependent fingerprint for the combined SDFT spectrometer within the bandwidth of the DFT spectrometer. When reconstructing a broad spectrum using such a calibration matrix, the speckle contribution of the MMW is minimal--as the algorithm averages it out as noise--thus allowing one to accurately reconstruct a broad spectrum. On the other hand, when the reconstruction is performed within a smaller spectral window below the resolution limit of the DFT-only device, the output of the MZIs acts as a slowly varying DC-like offset, while the contribution of the speckle pattern on the calibration matrix becomes significant. This behavior allows the algorithm to robustly reconstruct both a broad and high-resolution spectrum and circumvent the resolution-bandwidth tradeoff.   

With the knowledge of such a calibration matrix, the spectral content of an unknown light input to the spectrometer can be reconstructed by solving $S=A^+ P_{out}$, where $A^+$ is the pseudoinverse of the SDFT calibration matrix A. If the number of measurements is smaller than the wavelength points to be reconstructed, the system of linear equations is under-constrained. Such constrained linear equations can be solved using least square minimization \cite{hansen_1990}, such as the elastic-net regularization technique \cite{kita_2018},
\begin{equation}
 S(\lambda)\equiv \argmin_{S,S>0} \left\{||P_{out}-A.S||^2 + l_1  ||S||_1 + l_2 ||S||_2^2\right\}
\label{eq:refname4aa}
\end{equation}

where $l_1$ and $l_2$ are regularization hyperparameters that are appropriately selected depending on the density of the reconstructed spectrum. $l_2=0$ gives a well-known lasso regularization used for compressive sensing on sparse signal, and $l_1=0$ gives a 2-norm Tikhonov regularization (or ridge regression) appropriate for reconstructing a dense signal. Such regularization technique allows a robust reconstruction of the input signal over noisy or unconstrained data.  

\subsection{Statistical analysis of the device}
\begin{figure}[t]
\centering
\includegraphics[width=12cm]{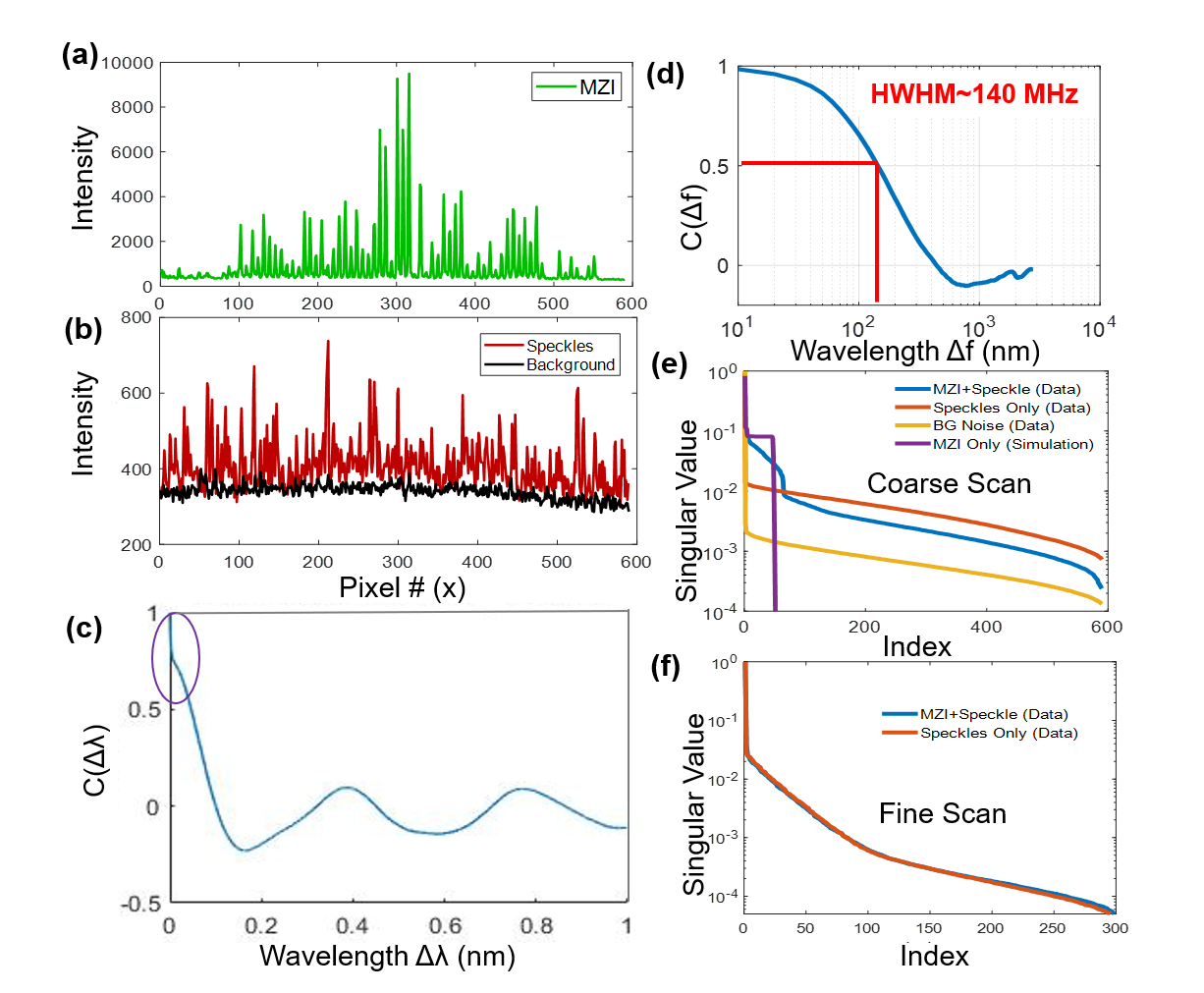}
\caption{ A 1D array of the intensity values recorded (a) at the output of the SDFT and (b) speckle-only region. The black curve in (b) shows the contribution of background along with detector noise. (c) Normalized spectral correlation averaged over all pixels generated from the SDFT region of the spectrometer consisting of both MZI and speckle output. The narrow peak circled in the plot is due to the highly de-correlated speckle pattern and the slowly varying feature is due to the DFT spectrometer. (d) Averaged intensity correlation data generated from a separate high-resolution RF-scan measured by stepping the RF frequency with 10 MHz step. The decorrelation width of the SDFT spectrometer is 140 MHz, which sets the true resolution of the device. (e-f) Comparison of the singular values for MZI-only (theory), speckle-only (exp), SDFT spectrometer (exp) and the background noise (exp). The comparison is shown for (e) coarse scan over a broad wavelength window and (f) high-resolution fine scan over a small bandwidth.}
\label{fig:figure3}
\end{figure}

The performance of the SDFT spectrometers and the information content available for reconstructing the spectrum can be studied by performing statistical and linear algebra analyses of the images recorded at the output of the chip. 1D intensity values recorded from the SDFT region and speckle-only region are plotted in Fig. \ref{fig:figure3} (a) and (b). In this measurement, the relative contribution of the speckle on the MZI array is $\sim$15\%. The speckle contribution to the SDFT device can be increased by intentionally misaligning the input fiber such that more light passes through the substrate. The black curve shows the contribution of the background and detector noise.

The spectral resolution of the device is set by the wavelength correlation of the output intensity pattern and is given by \cite{rawson_1980, redding_2013},

\begin{equation}
C(\Delta \lambda,x)=\frac{\left\langle I(\lambda,x)I(\lambda+\Delta \lambda,x)\right\rangle}{\left\langle  I(\lambda,x)\right\rangle \left\langle I(\lambda+\Delta \lambda,x)\right\rangle} -1,
\label{eq:refname6}
\end{equation}

where  $\left\langle I(\lambda,x)\right\rangle$ is the intensity recorded by the camera at position $x$ for an input optical wavelength $\lambda$ averaged over all spatial positions. The spectral resolution ($\delta \lambda$) of the device is set by the correlation width at which the speckle correlation drops to half. 

Figure \ref{fig:figure3} (c) is the normalized wavelength correlation of the SDFT spectrometer measured by averaging over multiple pixels recorded at the output of the chip. The data shows two distinct features, a rapid intensity de-correlation is overlaid on top of a slow de-correlation. The narrow peak circled in the plot is due to the contribution of the rapidly de-correlating effect of speckle and the slowly varying feature is due to the DFT spectrometer. To further probe the fast de-correlation behavior of the device, we performed a high-resolution spectral scan using a radio-frequency (RF) scan technique developed by Scofield et al. \cite{scofield_2018}. A CW laser is modulated to suppress the carrier frequency and the sidebands are scanned in 10 MHz steps using a computer controlled RF driver. This results in a new calibration matrix that is the sum of two mirror wavelengths that are detuned equally from the carrier frequency. Using this technique, a frequency dependent calibration matrix is recorded where the SDFT matrix is constructed by summing rows of pixels from the SDFT region. The high-resolution normalized correlation matrix measured from the device is plotted in Fig. \ref{fig:figure3} (d). From the data, we obtain the resolution of the combined SDFT device to be 140 MHz, $\sim 160$ times better than the DFT-only spectrometer. This unequivocally demonstrates that the addition of speckle to the DFT spectrometer, thus forming a SDFT spectrometer, enhances the resolving power of the device. The resolution of the MMW spectrometer can be estimated as $\delta\lambda \sim (\lambda/n)^2/(2nL(1-cos(NA)))$ \cite{redding_2013} and the measured resolution is consistent with the predicted resolution for an 8.5 mm long MMW within the uncertainties in the properties of the waveguide. A more detailed comparison of the 2D spatial and frequency correlations along with the impact of length of MMW and the device structure on the decorrelation width is given in the Appendix (B-D). 

Further insights on the information content carried by the SDFT matrix can be gained by performing singular value decomposition (SVD) analysis of the calibration matrix \cite{piels_2017, hansen_1990,wang_2014}. A rectangular calibration matrix ($A_{SDFT}\in \mathbf{R}^{n\times m}$ ) built by scanning $m$ wavelength steps and summing 2D camera pixels column wise to generate $n$ measurements can be decomposed as $A_{SDFT}=U\Sigma V^T$. $U$ and $V$ are left- and right-eigenbasis of the matrix and $\Sigma$ is an $m\times n$ diagonal matrix containing $r$ non-zero singular values, where $r$ is the rank of the calibration matrix and corresponds to the uncorrelated eigenvectors for signal reconstruction. Singular values are square roots of the eigenvalues of the $A_{SDFT}^TA_{SDFT}$ matrix and are arranged in descending order. The larger singular values capture most of the signal information contained in the calibration matrix and the values closer to zero simply add noise to the system. 

To perform comparative studies between speckle-only, DFT-only, and SDFT spectrometers, we generated calibration matrices for each device by summing rows of pixels from the SDFT region and speckle-only region as shown in Fig. \ref{fig:figure2} (a). The calibration matrices are normalized to have a unit Frobenius norm so that the relative magnitude of the singular values can be compared \cite{piels_2017}. The DFT-only calibration matrix is numerically simulated \cite{florjanczyk_2007}. Figure \ref{fig:figure3} (e) is the comparison of singular values for various devices at a wider bandwidth regime (10 nm with 0.01 nm steps). A large number of distinctive eigenfunctions with larger values allows for better signal reconstruction \cite{wang_2014, piels_2017, haldrup_2014}. The SDFT spectrometer consists of 64 large singular values corresponding to the 64 MZI channels that provide coarse resolution and the additional $\sim 500$ smaller eigenvalues corresponding to high-frequency eigenfunctions that provides enhancement in spectral resolution. As the data indicates in Fig. 3 (e), the addition of speckle extends the number of eigenvectors available for signal reconstruction (blue curve) for the SDFT spectrometer over the rank deficient DFT-only spectrometer (purple curve). The extended singular values are above the background noise level (yellow). This added contribution results in the increase in the resolution of the device over the traditional DFT spectrometer and bandwidth increase over the speckle-only spectrometer \cite{piels_2017}. 

Figure \ref{fig:figure3} (f) is the comparison of singular values calculated from a calibration matrix recorded over a smaller wavelength region (50 pm) but with a finer spectral scan steps (0.08 pm). As can be seen, the speckle calibration matrix has near identical property as the SDFT spectrometer. This is a result of the contribution from MZI array changing very little within such a narrow wavelength scan window thus demonstrating that the addition of the DFT does not degrade the functionality of the MMW speckle. This dual feature indicates that combining coarse resolution, higher bandwidth DFT and high resolution, low bandwidth speckle spectrometer enables one to perform both high-resolution and broad-bandwidth spectra reconstruction using a single device. In this study the input signals were reconstructed using 2D camera output projected to form a 1D array of up to 640 elements. However, for an input signal with a dense spectral content, utilizing the entire 2D pixel array from \ref{fig:figure2} (a) ($\sim 150 \times 640$) provides $\sim$ 96,000 spectral channels instead of only 640 channels in the 1D matrix. This additional intensity information should result in a more robust signal reconstruction for a dense input signal.           

In order for the SDFT spectrometer to provide a unique high-resolution fingerprint over a wide wavelength range, the resolution of the MZI needs to be commensurate with the free spectral range of the MMW. To achieve both high-resolution and broad-bandwidth reconstruction, first a coarse spectrum limited by the resolution of the DFT can be reconstructed using a larger regularization parameter. Once the coarse spectrum is identified, the high-resolution spectrum can be reconstructed using a subset of the calibration matrix limited within the narrow spectral bandwidth using the above mentioned signal processing techniques. See Appendix E for example reconstruction simulated for the SDFT spectrometer. A single high-resolution broad bandwidth calibration matrix and a single measurement is sufficient for signal reconstruction. The bandwidth of this hybrid device is estimated to be $B \sim N/2 \times \Delta\lambda$, where $\Delta \lambda$ is the bandwidth of the speckle spectrometer and N is the number of MZIs. 
\section{Spectrum reconstruction}

\begin{figure}[t]
\centering
\includegraphics[width=12cm]{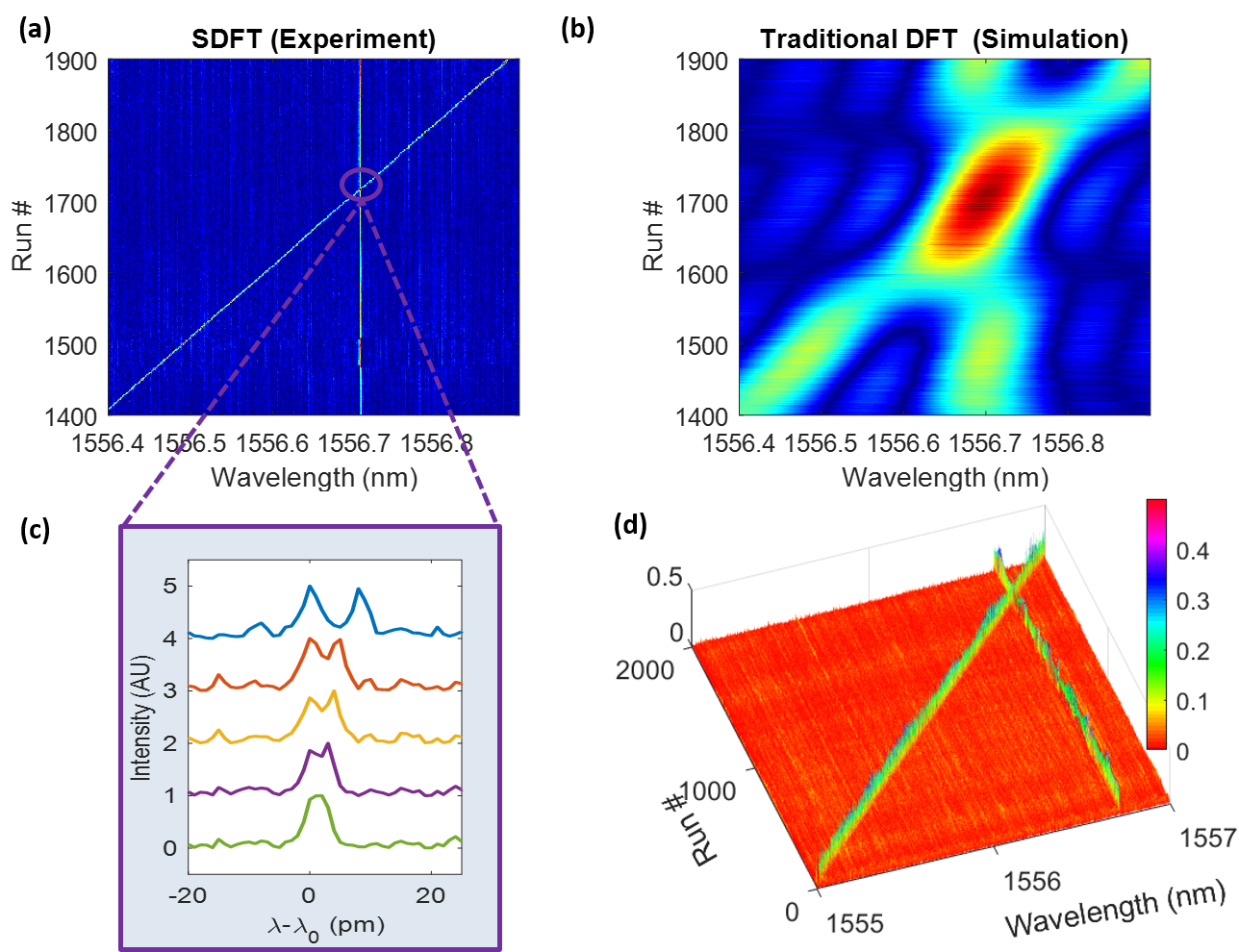}
\caption{(a) Two-tone spectral reconstruction test measured using the hybrid SDFT spectrometer where two peaks separated by 3 pm are resolved by the device. (b) The theoretical simulation of the performance of a comparable DFT-only spectrometer without the addition of speckle. The resolution of the DFT-only device is 182 picometer. (c) The zoomed-in plots show the two peaks separated by 2.9 pm are resolved. (d) Spectral reconstruction performed for 2000 unknown spectral configurations that are within a 2 nm window.}
\label{fig:figure4b}
\end{figure}

To demonstrate the functionality of the device, a series of two-tone reconstructions are performed using two monochromatic lasers. Two-tone tests were performed by simultaneously sending one fixed and one tunable laser through the spectrometer. A series of intensity patterns at the output of the SDFT region were recorded while scanning the relative detuning between the two input lasers. The laser is scanned at 1 pm step from 1555 nm to 1557 nm. Using the calibration matrix and the recorded output intensity pattern, we are able to reconstruct the dual wavelength input spectrum as the tunable laser is stepped across the entire range. A series of reconstructed spectra using the intensity pattern recorded from the SDFT device is plotted in Fig. \ref{fig:figure4b} (a) where the y-axis corresponds to different measurements with the relative detuning between the two input lasers changed. The vertical line represents the fixed wavelength laser and the diagonal line represents the tunable laser. The data is reconstructed using $l2$ regularization with a small regularization parameter, where the weight of the regularization is directly related to the reconstruction resolution \cite{hansen_1994}. A detailed analysis on the signal reconstruction technique and the effect of regularization on computational spectrometers are given in Refs. \cite{wang_2014, kita_2018}.

To compare the performance of the device, we theoretically simulated a calibration matrix and intensity pattern generated by the DFT-only spectrometer with comparable MZI parameters. The reconstructed spectrum is plotted in Fig. \ref{fig:figure4b} (b) for comparison. As the figure demonstrates, the SDFT spectrometer far outperforms the resolving capacity of the DFT-only spectrometers and is able to resolve two closely spaced spectra. Figure \ref{fig:figure4b}(c) is a zoomed-in reconstruction of the spectrometer where two lines separated by 3 pm (374 MHz) are resolved by the SDFT spectrometer, far beyond the resolution limit (182 pm, 22.7 GHz) expected from the DFT-only spectrometer. The measured resolution is limited by the uncertainty in the absolute wavelength accuracy of the laser, which is $\sim$ 2.5 pm. The high-resolution spectral correlation data in Fig.\ref{fig:figure3}(d) is not affected by this uncertainty, as the resolution is set by the jitter in the RF source and the laser linewidth, which are both below the measured 140 MHz spectral resolution. The reconstruction was performed for 2000 spectral steps of the tunable laser at 1 pm per step within a 2 nm (250 GHz) window, plotted in Fig. \ref{fig:figure4b} (d). The bandwidth far exceeds the bandwidth of the speckle-only spectrometer with comparable resolution \cite{redding_2014}. 

\begin{figure}[ht]
\centering
\includegraphics[width=10cm]{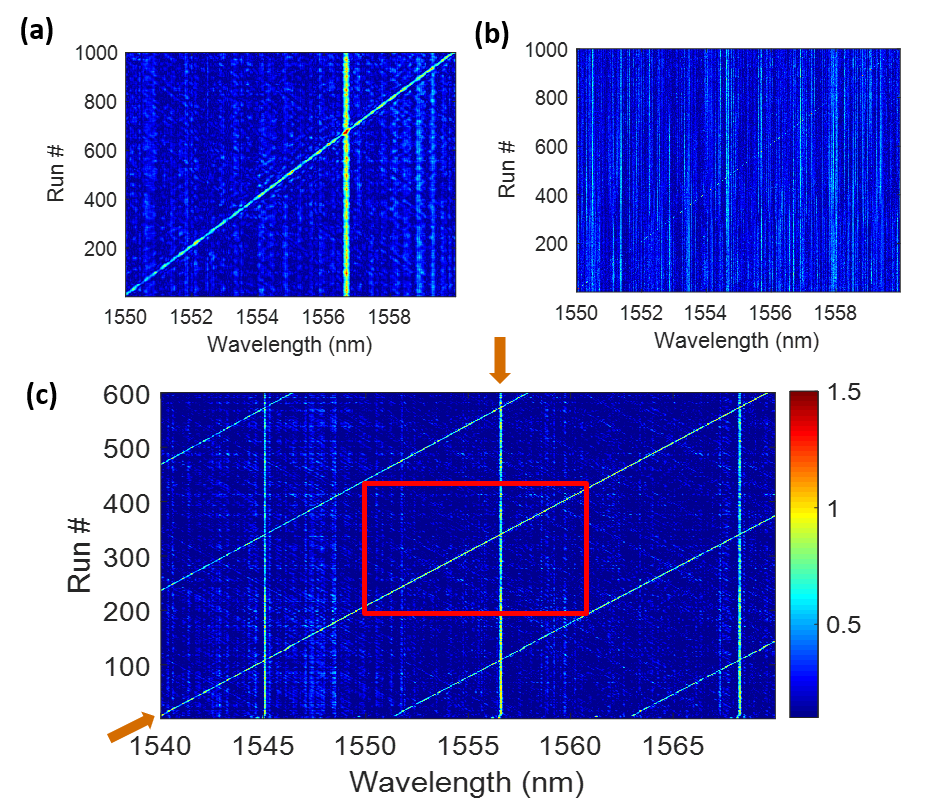}
\caption{ Two-tone spectrum reconstructed over a 10 nm window using (a) the hybrid SDFT and (b) speckle-only device. The speckle data is recorded from below the MZIs array where there is minimal to zero overlap from the MZIs output. As the spectrum indicates, the hybrid SDFT device is able to reconstruct the spectrum, whereas the speckle-only device reconstructs multiple false spectral contents. This is due to the limited unique speckle patterns as the input spectral window is increased. (c) Two-tone spectra reconstructed over a 30 nm window where one laser is fixed near 1556 nm and the second laser is scanned from 1540 nm to 1570 nm. Due to the limited bandwidth, or free spectral range, of the DFT spectrometer the algorithm reconstructs multiple false lines that are separated by 12 nm from the true spectrum, corresponding to the free-spectral-range of the DFT component of the SDFT spectrometer. The orange arrows indicate the location of the true input spectra. The red box is a 12 nm window where only the two spectra are present. This corresponds to the true bandwidth of the device.}
\label{fig:figure5}
\end{figure}

To experimentally demonstrate the spectral range of the SDFT device, we repeated the two-tone experiment over 10 and 30 nm windows. The calibration matrix of the SDFT device is recorded by summing the pixel arrays from the dashed white region in Fig. \ref{fig:figure2}(a), where the output is an overlap of MZIs and speckle modes. To compare the performance of the device with the speckle-only spectrometer, the speckle calibration matrix is generated by recording pixels below the MZIs array (dashed red region in Fig. \ref{fig:figure2}(a)) where the multimode speckle data has minimal to no contribution from the MZI output. The two-tone reconstruction experiment is repeated for both systems, where the data is acquired simultaneously within the same shot of the measurement. The reconstructed spectra using the SDFT and speckle-only spectrometers are plotted in Fig. \ref{fig:figure5} (a) and (b). As can be seen, the SDFT device is able to reconstruct the spectrum for a larger bandwidth region, whereas the speckle-only reconstructs multiple false spectra in addition to the true spectra. This is due to the limited number of unique speckle fingerprints available as the input spectral window is increased. A detailed theoretical comparison of the performance of the two devices is given in Appendix E. 
   
To display the spectral range of the SDFT, we collect data over a 30 nm sweep of the tunable laser. As can be seen in Fig. \ref{fig:figure5} (c), the reconstruction pattern repeats after 12 nm, thus indicating the bandwidth of the device. The arrows indicate the true spectral location of the input lasers with the enclosed red box consisting of true spectra within the free-spectral range of the DFT spectrometer (12 nm). The measured bandwidth is twice the predicted bandwidth for a DFT spectrometer \cite{velasco_2013}. This additional enhancement in bandwidth is attributed to the combination of the speckle and the reconstruction algorithm \cite{piels_2017}. 

All the data reported in this Article are taken in an open lab setting without precise temperature stabilization of the chip. The test data is typically taken within 30 minutes of the calibration. The device has been tested over an input power range of 1 to 6 mW. The dynamic range of the reconstructed spectrum is measured to be $\sim 12$ dB, which is limited by the measured signal-to-noise (SNR) of the output intensity and noise induced by system instability. This should be partly mitigated by packaging the device to thermally and mechanically isolate it from external perturbations and the measured signal can be increased by coating the unused surfaces of the chip with high-reflecting mirrors.

Speckle generated by multimode fibers are known to be particularly sensitive to small strain or temperature fluctuations \cite{redding_2013}. Detailed analyses of the effect of temperature drift of such speckle and DFT spectrometers are reported in Refs. \cite{redding_2014, reddingnoise_2014, herrero_2017}. The small footprint of the device ($\sim 1 cm^2$) with a shorter path length partially mitigates the thermal and strain issues that a multimode fiber speckle spectrometer suffers from. In addition, it has been reported that the speckle patterns generated by input light with wavelength $\lambda$ at temperature $T+\delta T$ and by input light at wavelength $\lambda+\delta\lambda$ at temperature $T$ are the same \cite{redding_2014}. Thus any spectral drift due to a small change in temperature can be compensated by a single correction offset in the reconstructed spectrum. Alternatively, the temperature of the spectrometer can be monitored and controlled by fabricating a layer of metallic heater on top of the device. 

\section{Conclusion}
In this Article, we demonstrate a novel chip-scale, passive spectrometer that combines a discrete Fourier-transform and a speckle functionality in a single device to significantly increase the finesse over the individual DFT and speckle-only spectrometers. We demonstrate that the device can resolve two lasers separated by 3 pm and we determine the true resolution of the device to be $\sim 1 $ pm (140 MHz) using intensity correlation measurements. The device has $\sim 12 $ nm bandwidth within an operational window of 100 nm in the 1500 nm-1600 nm region. The finesse of our device is $\sim 10,000$, two orders of magnitude larger than individual DFT or speckle spectrometers \cite{florjanczyk_2007, okamoto_2010, velasco_2013, nedeljkovic_2016, kita_2018,redding_2013,reddingNatPhot_2013}. To achieve the experimentally demonstrated bandwidth and resolution reported in this Article using a DFT-only spectrometer \cite{florjanczyk_2007, okamoto_2010} would require 10,000s of MZIs, making it infeasible for a chip-scale device. Even though speckle-based chip-scale spectrometers can achieve such high resolution, they are severely limited to a small operational bandwidth (0.1 nm) \cite{redding_2014}. The competing technologies involve using two different devices with coarse and fine resolutions \cite{redding_2013} or using optical switches or spectral multiplexing \cite{liew_2016} to achieve high-resolution and large-bandwidth reconstruction simultaneously. This significant improvement in the finesse was achieved even though the footprint and bandwidth of our device was not optimized. This device can be trivially extended in the same platform to have $>100$ nm bandwidth centered at $\sim 1550$ nm by fabricating 128 MZIs with $\Delta L_{min}  \leq 1.9 \mu$m. The operational range of the device is set by the wavelength specifications of the Y-splitter and edge coupler but can be extended to anywhere within the transparency window of silicon. In addition, by using heterostructures of different materials, the design can be extended to operate at a much wider range of wavelengths of interest.  

\appendix

\section*{Appendix A: Wavelength dependent raw output of the SDFT spectrometer}
 See Visualization 1 for  wavelength dependent raw output of the SDFT spectrometer. Even though in the text $\sim 40$ pixel arrays were summed to obtain a calibration matrix, a calibration matrix constructed with a single pixel array consisting of both MZI and speckle output was sufficient to reconstruct both high-resolution and broad bandwidth spectra from the 2-tone tests. As can be seen in the video, speckle generated from the multimode wafer changes rapidly with the change in the input wavelength by a picometer. 

\section*{Appendix B: Spatial and spectral correlation comparison between the speckle and SDFT spectrometers}
\begin{figure}[htbp]
\centering
\includegraphics[width=10cm]{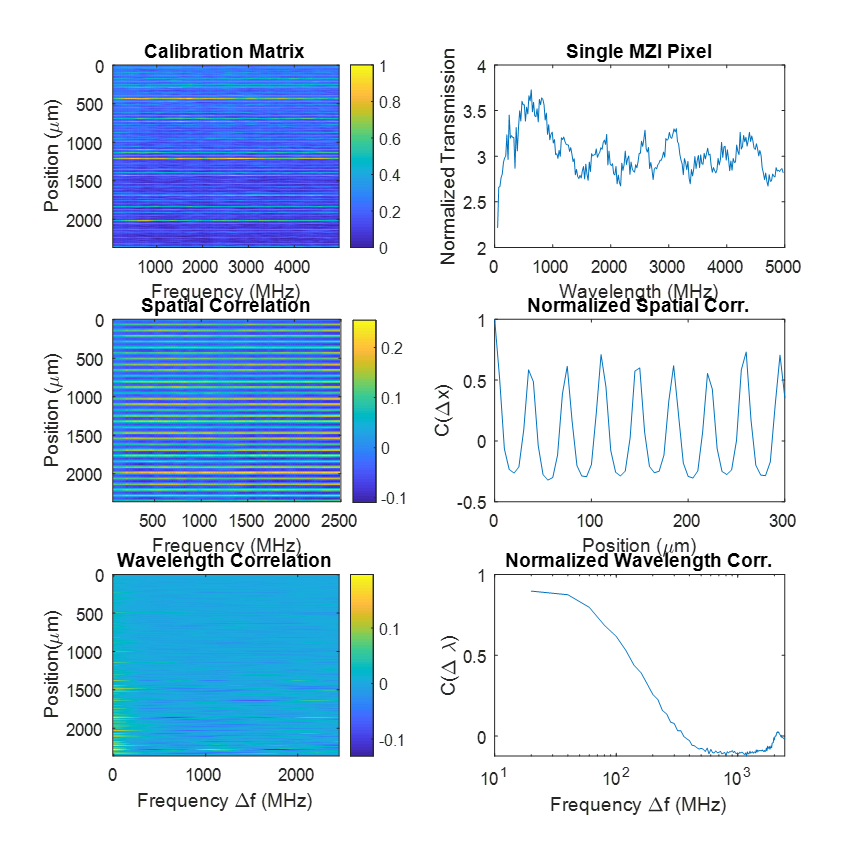}
\caption{Spatial and spectral correlation of the SDFT spectrometer where the intensity consists of both MZI array output and speckle.}
\label{fig:figure2Sa}
\end{figure}

\begin{figure}[h!]
\centering
\includegraphics[width=10cm]{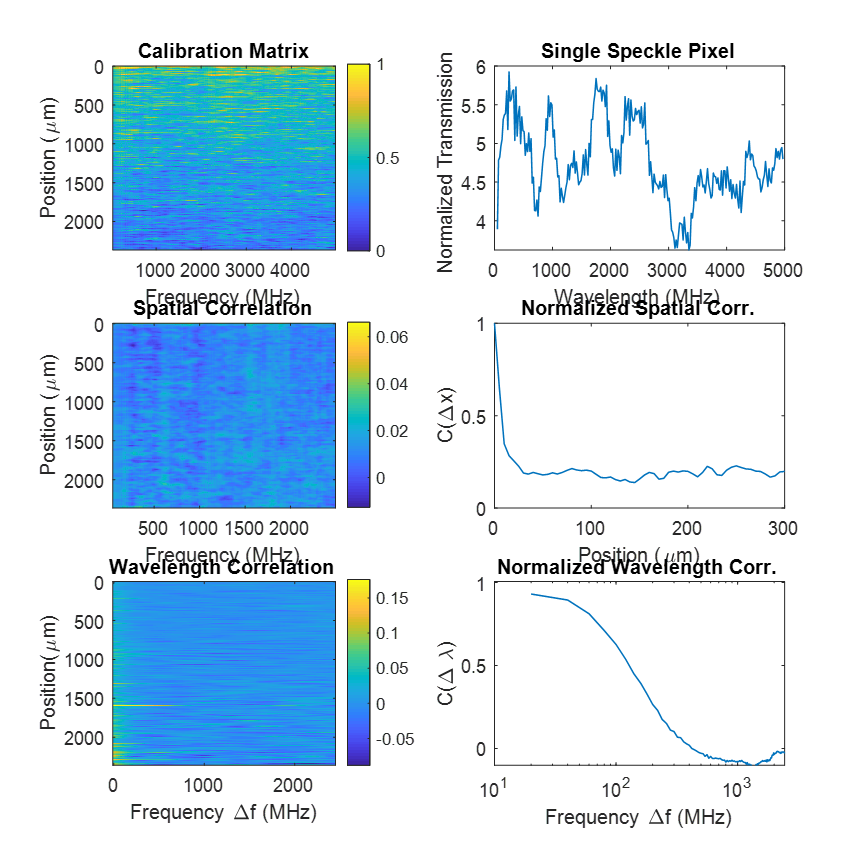}
\caption{Spatial and spectral correlation of the speckle-only spectrometer taken from the region of the chip without the MZI array.}
\label{fig:figure2Sb}
\end{figure}

Fig. \ref{fig:figure2Sa}  and Fig. \ref{fig:figure2Sb} show the transmission matrix and its 2D spectral and spatial correlation for SDFT and speckle-only spectrometers. Both calibration matrices are recorded simultaneously from a high-resolution RF-scan. The spatial correlation width for the MZI is 5 $\mu$m, which is comparable to the optical mode at the output of the inverse taper structure used for terminating the waveguides. The repeating peaks correspond to the spatial separation of the MZIs' output separated by $40 \mu$m. The speckle lobes are comparable to the MZI modes. The spectral correlation width is comparable for both spectrometers ($< 140$MHz) as they are primarily due to the speckle decorrelation. 

\section*{Appendix C: Spectral resolution of the speckle spectrometer as a function of the wafer length}
\begin{figure}[htbp]
\centering
\includegraphics[width=10cm]{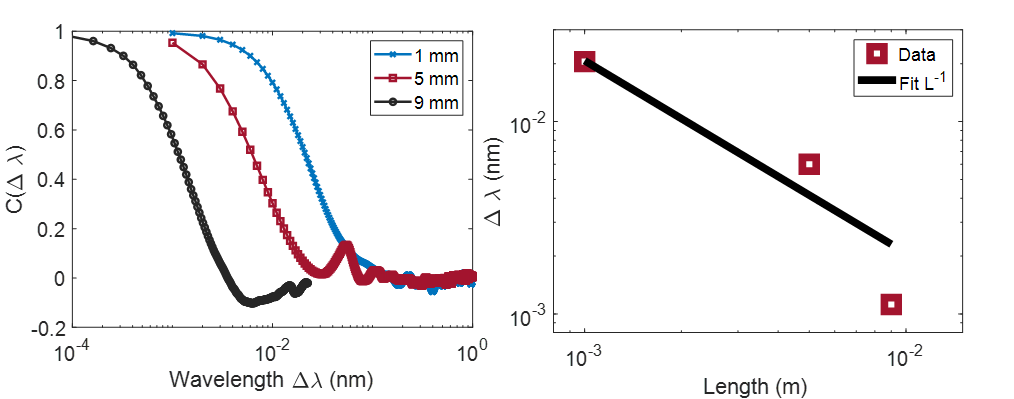}
\caption{ (a) Spectral correlation of the speckle spectrometer for three different wafer lengths. As the length of the wafer is decreased, the spectral correlation width increases. (b) The correlation width is fitted with an inverse length as predicted by \cite{redding_2013}.}
\label{fig:figure3S}
\end{figure}

Fig. \ref{fig:figure3S} shows the spectral resolution of the speckle spectrometer as a function of the wafer length. As the length of the wafer is decreased, the spectral correlation width increases. The correlation width is fitted with an inverse length as predicted by \cite{redding_2013}. The test wafer consists of 675 $\mu$m thick silicon with a $2 \mu$m oxide layer on top. The test wafers were 5 mm wide. With a 1 mm long silicon wafer, we are able to achieve $\sim 20$ pm spectral resolution.

\section*{Appendix D: Influence of the thin silicon device layer on the spectral resolution of the speckle spectrometer}
\begin{figure}[htbp]
\centering
\includegraphics[width=6cm]{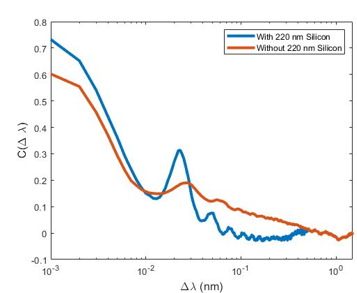}
\caption{ (a) Spectral correlation of the speckle spectrometer for a device with and without 220 nm silicon layer. The presence of the silicon layer has minimal to no impact on the resolution of the speckle spectrometer.}
\label{fig:figure4S}
\end{figure}

The typical wafer used for fabricating MZIs includes a thin 220 nm silicon layer, on top of the oxide layer, in which the photonic devices are etched. Figure \ref{fig:figure4S} shows spectral correlation of a speckle spectrometer built using a silicon wafer with and without a 220 nm silicon layer on top of the oxide layer. The presence of the silicon layer shows minimal to no impact on the resolution of the speckle spectrometer.
 
\section*{Appendix E: Theoretical comparison of SDFT vs speckle-only vs DFT-only spectrometer}
To demonstrate the dual functionality of a hybrid speckle enhanced DFT spectrometer, we simulated the output of a SDFT spectrometer by combining the intensity output of the speckle and DFT waveguides. The performance of the simulated SDFT spectrometer is compared with the DFT-only and speckle-only spectrometers. The DFT spectrometer is simulated using 64 MZIs with the relative path length difference increasing in an increment of 50 $\mu$m with the index of refraction to match with single mode 500 nm $\times$ 220 nm silicon waveguides \cite{florjanczyk_2007}. The interferometer outputs are mapped to a 1D array of camera pixels. Similarly, the speckle spectrometer is simulated using techniques described in \cite{valley_2016, paudel_2020} using a 500 $\mu$m wide slab silicon waveguide. The spectral correlation width of the speckle waveguide is simulated to be $\sim 150$ MHz in order to mimic the experimentally obtained data. 

\begin{figure}[htbp]
\centering
\includegraphics[width=10cm]{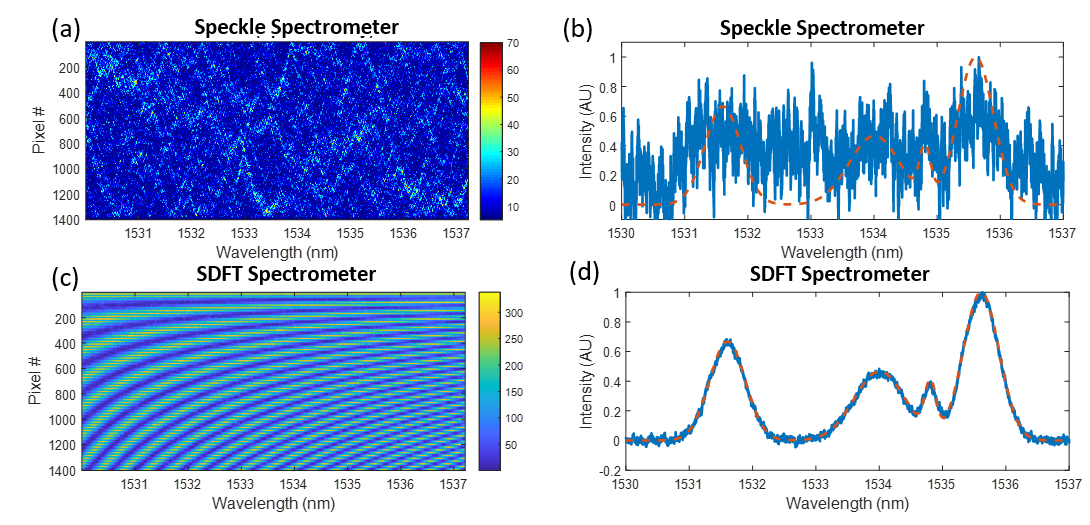}
\caption{ (a)  Calibration matrix of the simulated high-resolution speckle-only spectrometer. As can be seen in the image, speckle patterns start repeating as the wavelength is scanned for a wider bandwidth. (b) The orange curve is the broadband input spectrum and the blue is the reconstructed spectrum by the speckle-only device. The speckle-only device fails to reconstruct the broad input spectrum. (c) Calibration matrix of the combined SDFT device and (d) the reconstructed broadband spectrum using the SDFT device is plotted in blue.}  
\label{fig:figure6S}
\end{figure}

A wavelength dependent calibration matrix is generated by adding the DFT intensity with 10\% of the speckle intensity. The simulated calibration matrix for the SDFT spectrometer is plotted in Fig. \ref{fig:figure6S}. 

The limitation of the high-resolution speckle spectrometer for reconstructing broad spectra is illustrated in Fig. \ref{fig:figure6S}. The output power of the spectrometers ($P_{out}$) is simulated by transforming a randomly generated input spectral content (S) with the calibration matrix (A) following the techniques used in \cite{redding_2013, piels_2017}. Figures \ref{fig:figure6S} (a) and \ref{fig:figure6S} (c) are the simulated calibration matrices of the speckle spectrometer and the SDFT device respectively. As the number of unique speckle patterns is proportional to the number of optical modes sustained in a multimode waveguide, the high-resolution speckle data can provide a unique fingerprint only for a limited bandwidth (B), beyond which the speckle pattern repeats with the change in input wavelength. This results in poor spectral reconstruction when using the speckle-only data for reconstructing coarse features beyond its bandwidth.  

The reconstruction limitation of such a high-resolution device for broadband spectra is illustrated in Fig. \ref{fig:figure6S} (b), where a broad input spectrum (orange curve) composed of various Gaussian peaks is input to the spectrometer. The blue curve is the reconstructed spectrum calculated using the calibration matrix plotted in Fig. \ref{fig:figure6S} (a). As evident from the figure, the speckle-only device fails to reconstruct the broad input spectrum. To contrast its performance with the SDFT device, we reconstruct the same input spectrum using the SDFT calibration matrix generated by combining the speckle and DFT data. The broadband spectrum reconstructed by the SDFT device is plotted as the blue curve in Fig. \ref{fig:figure6S} (d) overlapped with the input orange spectrum. The result demonstrates that the SDFT device can reconstruct a broadband spectrum despite the presence of high-frequency intensity perturbations on the DFT transmission matrix due to the speckle. 

\begin{figure}[t]
\centering
\includegraphics[width=10cm]{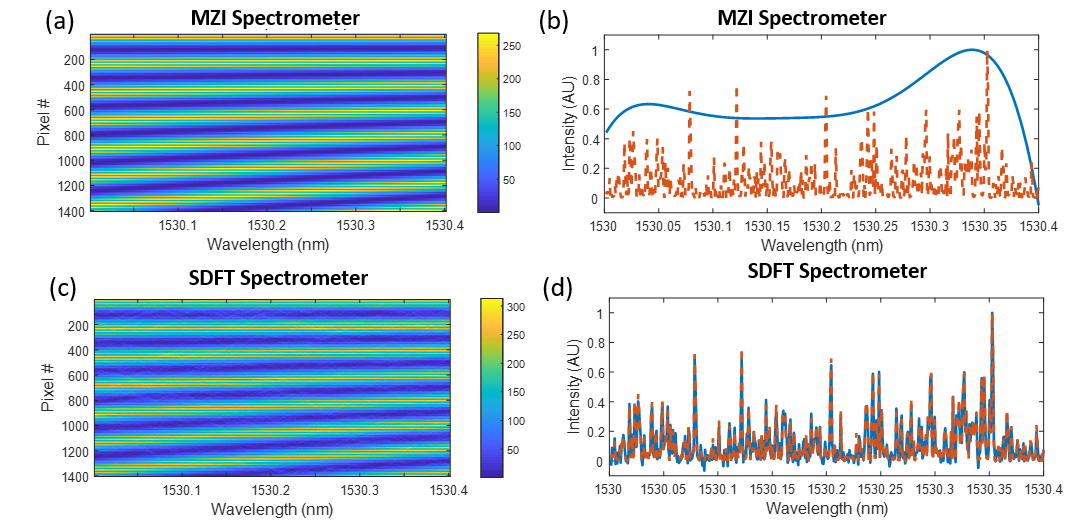}
\caption{ Zoomed-in calibration matrix of (a) a DFT-only spectrometer and (c) the SDFT spectrometer from Fig. \ref{fig:figure6S}. (b) The orange curve shows dense, closely spaced multiple Gaussians that are input to the DFT spectrometer. The blue curve is the reconstructed spectrum, demonstrating that the broadband DFT-only spectrometer is incapable of reconstructing high-resolution spectra.  (d) shows the reconstructed spectrum for the same input data using the SDFT device. The result demonstrates that the SDFT device is still capable of resolving a high-resolution, densely-spaced spectrum even when only a small fraction (10\%) of speckle intensity is added to the MZI calibration matrix.}  
\label{fig:figure6Sb}
\end{figure}

To demonstrate that the same combined spectrometer is capable of reconstructing closely spaced, high-resolution spectra, we generated a new set of input specta S following Ref. \cite{piels_2017} and re-ran the reconstruction algorithm with a much narrower spectral window. Figure \ref{fig:figure6Sb} (a) is the calibration matrix for a DFT-only spectrometer. Figure \ref{fig:figure6Sb} (c) is the zoomed-in calibration matrices from Fig. \ref{fig:figure6S} (c)  corresponding to the hybrid SDFT spectrometer. In contrast to the case described in Fig. \ref{fig:figure6S}, here the hybrid SDFT spectrometer is able to reconstruct the high-resolution spectra, whereas the DFT-only spectrometer fails to do so.  The dashed orange curves in Fig. \ref{fig:figure6Sb} (b) and Fig. \ref{fig:figure6Sb} (d) show densely spaced, multiple Gaussian peaks with varying intensities input to the DFT and hybrid SDFT spectrometers. As expected, Fig. \ref{fig:figure6Sb} (b) shows the DFT-only device fails to reconstruct the high-resolution spectra, as the spectral content is beyond the resolving limit of the device. As can be seen in Fig. \ref{fig:figure6Sb} (d), the hybrid SDFT spectrometer is able to reconstruct the densely spaced, high-resolution spectra very well. 

The reconstructions shown in Fig \ref{fig:figure6S} (d) and \ref{fig:figure6Sb} (d) demonstrate that the hybrid SDFT device can resolve both broadband and high-resolution spectra that otherwise cannot be reconstructed with either a DFT-only or speckle-only spectrometer alone.  

\section*{Funding Information}
U.S. Air Force (FA8802-19-C-0001)

\section*{Disclosures}
UP, TSR: The Aerospace Corporation (P)

\section*{Acknowledgments}

The authors would like to thank Dr. George C. Valley for providing Mathematica code to calculate waveguide speckle, Cameron Horvath for providing test control silicon wafers, Dr. Stefan Preble for providing a high NA fiber, Dr. Marta Luengo-Kovac for numerous discussions about the content of the paper, and Dr. Derek Kita for sharing his results on DFT spectrometer.


\bibliography{bibil_optica}

\end{document}